\documentclass[sort&compress,final]{aipproc}

\layoutstyle{8x11double}
\def\cerenkov{$\check{\rm C}$erenkov~}

\def\PR#1#2#3{Phys. Rev. {\bf #1}, #2 (#3)}
\def\PRL#1#2#3{Phys. Rev. Lett. {\bf #1}, #2 (#3)}
\def\PL#1#2#3{Phys. Lett. {\bf #1}, #2 (#3)}
\def\NL#1#2#3{Nucl. Phys. {\bf #1}, #2 (#3)}
\def\NP#1#2#3{Nucl. Phys. {\bf #1}, #2 (#3)}

\def\PTP#1#2#3{Prog. Theor. Phys. {\bf #1}, #2 (#3)}
\def\EPJ#1#2#3{Eur. Phys. J. {\bf #1}, #2 (#3)}

\def\PRD#1#2#3{Phys. Rev. {\bf D#1},~#2 (#3)}

\def\PLB#1#2#3{Phys. Lett. {\bf B#1} (#2) #3}

\def\etal{{\it et al.}}

\def\simgt{\lower.5ex\hbox{$\; \buildrel > \over \sim \;$}}

\def\datm{\delta m^2_{{\rm atm}}}
\def\dsun{\delta m^2_{{\rm sol}}}

\def\satm{\sin^2 2 \theta_{{\rm atm}}}
\def\ssun{\sin^2 2 \theta_{{\rm sol}}}
\def\srct{\sin^2 2 \theta_{{\rm rct}}}

\def\dmns{\delta_{_{\rm MNS}}}

\def\dm#1{\delta m^2_{#1}}

\def\nn{\nonumber}
\begin{document}
\title{
Resolve the Neutrino Parameter Degeneracies with
the T2K Off-axis Beam and the Large Detector in Korea
\footnote{{%
Proceedings of PASCOS2005.
This talk is based on Ref\cite{T2K2}.
}}}
\author{Naotoshi Okamura
\thanks{okamura@yukawa.kyoto-u.ac.jp}
}
{
address={
Yukawa Institute for Theoretical Physics,
Kyoto University, Kyoto 606-8502, Japan.},
}

\begin{abstract}
 In this talk, 
we show the physics impacts of putting a large Water \cerenkov 
detector in Korea during the T2K experimental period. 
 The T2K experiment which will start in 2009 plans to use the high
intensity conventional neutrino beam from J-PARC at Tokai village, Japan.
 The center of this beam will reach the sea level between
Japan and Korea, and an off-axis beam at $0.5^{\circ}$ to
$1.0^{\circ}$ can be observed in Korea. 
 For a combination of the $3^{\circ}$ off-axis beam at SK with 
baseline length $L=295$km and the $0.5^{\circ}$ off-axis beam 
in the east coast of Korea, near Gyeongju,
at $L=1000$km, we find that the neutrino mass hierarchy
(the sign of the larger mass-squared difference)
can be resolved and the CP phase of the MNS unitary matrix can be
constrained uniquely at 3-$\sigma$ level when $\srct \simgt 0.06 $.
{
\vspace*{-60ex}
\begin{flushright}
 YITP-05-44, \,\,\, hep-ph/0509164 
\end{flushright}
\vspace*{+53ex}
}
\end{abstract}

\maketitle

The results of solar and atmospheric neutrino oscillation
experiments are consistent with the 3 neutrino model,
which has 6 observable parameters in neutrino oscillation experiments.
They are 2 mass-squared differences,
3 mixing angles,
and one CP phase.
The atmospheric neutrino oscillation experiments determine 
the absolute value of the larger mass-squared difference
($\datm=\dm{13}$)
and one mixing angle ($\theta_{\rm atm}$)
\cite{sk-atm04} as
\begin{eqnarray}
\label{atm-data}
1.5\times10^{-3} &<& 
\left|\dm{13}\right|
\equiv\left|m_{3}^2 - m_{1}^2\right|
< 3.4\times10^{-3} {\mbox{{eV}}}^2\,,\nn\\
0.92 &<&
\satm
\end{eqnarray}
at the 90\% confidence level.
The K2K experiment \cite{k2k04} confirms the above results.
The solar neutrino experiments \cite{kayser03}
and the KamLAND experiment \cite{kamland04} determine
the smaller mass-squared difference
($\dsun=\dm{12}$) and another mixing angle ($\theta_{\rm sol}$)
as
\begin{eqnarray}
\dm{12} &\equiv&
m_2^2 - m_1^2 = 8.2^{+0.6}_{-0.5} \times 10^{-5} {\mbox{{eV}}}^2
\,,\nn\\
 \tan^2\theta_{\rm sol} &=& 0.40^{+0.09}_{-0.07}\,.
\label{sol-data}
\end{eqnarray}
The CHOOZ reactor experiment \cite{chooz} gives 
the upper bound of the third mixing angle ($\theta_{\rm rct}$) as 
\begin{equation}
\srct < 0.16 \mbox{{~  for  ~}}
|\dm{13}| = 2.5\times10^{-3}\mbox{{eV}}^2\,,
\label{rct-data}
\end{equation}
at the 90\% confidence level.
The CP phase ($\dmns$) has not been constrained.
In the future neutrino oscillation experiments,
we should not only measure $\srct$ and $\dmns$, 
but also resolve the parameter degeneracies  
\cite{koike00,minakata01,barger01},
such as the sign of $\dm{13}$.
These parameters are related to the MNS \cite{MNS} matrix elements
as $\sin^2 \theta_{\rm rct} = |U_{e3}|^2$,
$\sin^2 \theta_{\rm atm} = |U_{\mu3}|^2$,
$\sin^2 2\theta_{\rm sol} = 4|U_{e1}U_{e2}|^2$.
The other elements are obtained from the unitary conditions
\cite{hagiwara-okamura,t2b}.

\begin{figure}
  \includegraphics[height=.19\textheight]{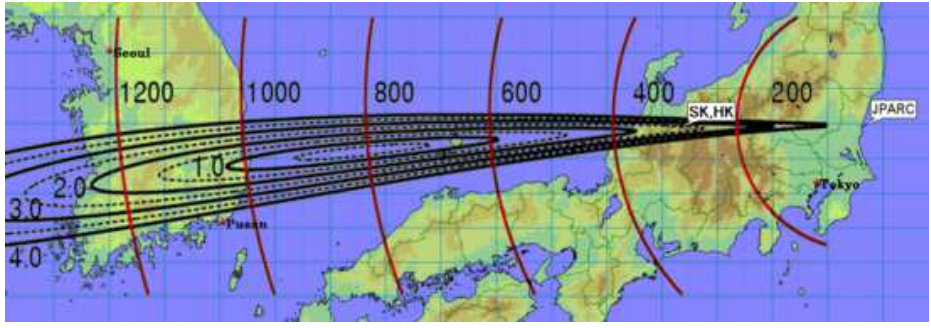}
\caption{
The off-axis angle of the neutrino beam from J-PARC on the sea level
when the beam center is $2.5^\circ$ off at SK.
The baseline length for $L=200, 400, 600, 800, 1000, 1200$km are shown
by vertical lines,
and closed curves stand for the off-axis angle between $0.5^\circ$
and $4.0^\circ$ with $0.5^\circ$ step.
}
\label{fig:mapL}
\end{figure}

\begin{figure}
  \includegraphics[height=.23\textheight]{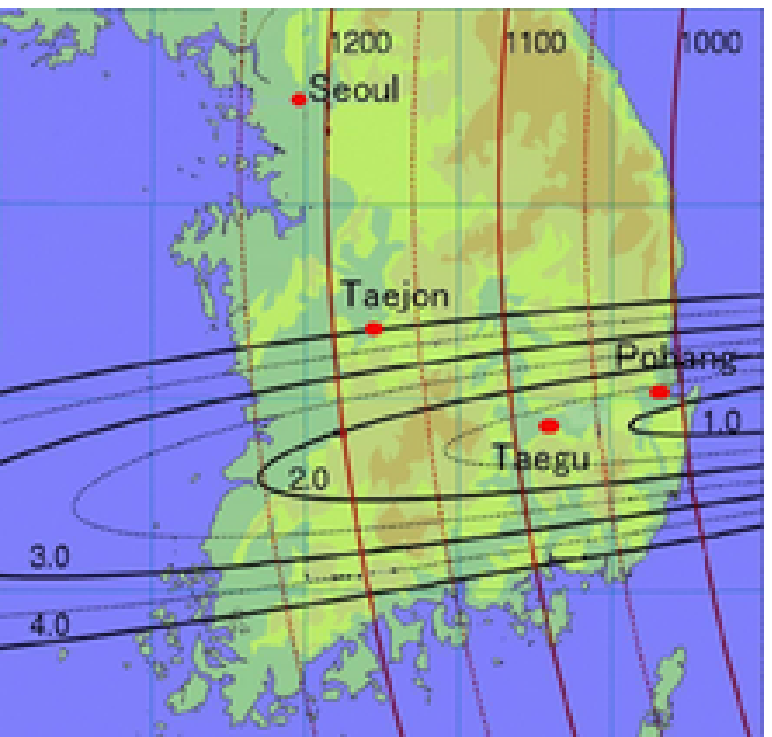}
~~~~~~~~~~~~~~~
  \includegraphics[height=.23\textheight]{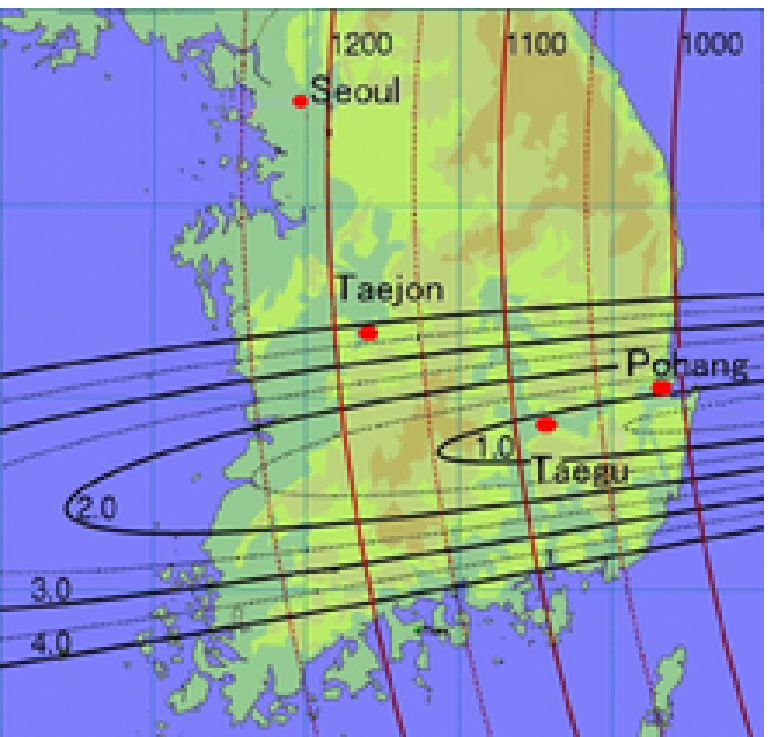}
\caption{
The magnified figures around Korea,
when the beam center is $2.5^{\circ}$(left) 
and $3.0^{\circ}$ (right) off at SK.  
The baseline length for $L = 1000, 1100, 1200$km are shown by 
vertical contours, and the off-axis angles are shown by elliptic 
contours between $0.5^{\circ}$ and $4.0^{\circ}$.
The interval of each line is $0.5^{\circ}$.}
\label{fig:map}
\end{figure}

In this talk,
we discuss the possibility of detecting in Korea
the neutrino beam from J-PARC (Japan Proton Accelerator Complex)
at Tokai village \cite{j-parc},
that will be available during the period of the T2K (Tokai-to-Kamioka) 
experiment \cite{t2k}.
In the T2K experiment,
the center of the neutrino beam will reach the sea level 
near the east coast of Korea.
At the baseline length $L = 295$km away from J-PARC,
the upper side of the beam at $2^\circ$ to $ 3^\circ$ off-axis angle
is observed at Super-Kamiokande (SK),whose fiducial volume is 22.5kt,
and
the lower side of the same beam at $0.5^\circ$ to $3.0^\circ$
appears in the east coast of Korea \cite{T2K2,hagiwara04}, 
at about $L = 1000$km; see Figs.\ref{fig:mapL},\ref{fig:map}. 
In order to quantify our investigation,
we study physics impacts of putting a
100kt-level water \cerenkov detector,
which allows us to distinguish clearly
$e^{\pm}$ events from $\mu^{\pm}$ events,
in Korea,
during the T2K experiment period, 
which is for 5 years with $10^{21}$ POT per year.
 As of today, September 2005, there is no proposal to construct a huge
water \cerenkov detector in Korea,
but we investigate the prospect of the neutrino oscillation experiment
with two detector putting the different baseline length.

We use the Charged-Current-Quasi-Elastic (CCQE) events in our analysis,
because they allow us to reconstruct the neutrino energy
by measuring the strength and the orientation of the \cerenkov lights.
Since the Fermi-motion effect of the target dominates 
the uncertainty of the neutrino energy reconstruction, about 80 MeV,
in the following analysis we take the width of the energy bin as 
$\delta E_\nu=200$~MeV for $E_\nu > 400$ MeV. 
The event numbers in the $i$-th energy bin,
$E_\nu^i < E_{\nu} <E_{\nu}^{i+1}$ where
$E_\nu^i = \delta E_\nu\times(i+1)$,
are then calculated as
\begin{equation}
\label{eq:N}
N_\beta^{i} (\nu_\alpha)=
 N
 \int_{E_\nu^i}^{E_\nu^{i+1}}
 \Phi_{\nu_\alpha}(E)~
 P_{\nu_\alpha\to\nu_\beta}(E)~ 
 \sigma_\beta^{QE}(E)~
 dE\,,
\end{equation}
where
$P_{\nu_{\alpha}\to\nu_{\beta}}$
is the neutrino oscillation probability including the matter effect, 
$N$ is the number of target nucleons,
$\Phi_{\nu_\alpha}$ is the $\nu_{\alpha}$ flux from J-PARC,
and
$\sigma_\beta^{QE}$ is the CCQE cross section of $\nu_\beta$
per nucleon in water.
We include the contribution from the secondary neutrino flux of the 
$\nu_\mu$ primary beam as the background events 
for the signal $e$- and $\mu$-like events.
After summing up the events from all flux,
the $e$-like and $\mu$-like events
for the $i$-th bin are obtained as
\begin{equation}
 N^i_\alpha = 
N_{\alpha}^{i}(\nu_\mu)+
N_{\alpha}^{i}(\nu_e) +
N_{\bar{\alpha}}^{i}(\bar{\nu}_e) +
N_{\bar{\alpha}}^{i}(\bar{\nu}_\mu)\,,
\end{equation}
where $\alpha=e$ and $\mu$.
In this analysis,
we do not add the background from $\tau$ pure leptonic decay
because the beam intensity above the $\tau$-lepton 
production threshold is small,
even for the $0.5^{\circ}$ off-axis beam
which has the hardest spectrum in our analysis.
We also do not add 
the events
from the Neutral Current interactions
and
from the Deep Inelastic Scattering
which induce the miss-identification for the signal event.

\begin{figure}
\includegraphics[height=.24\textheight]{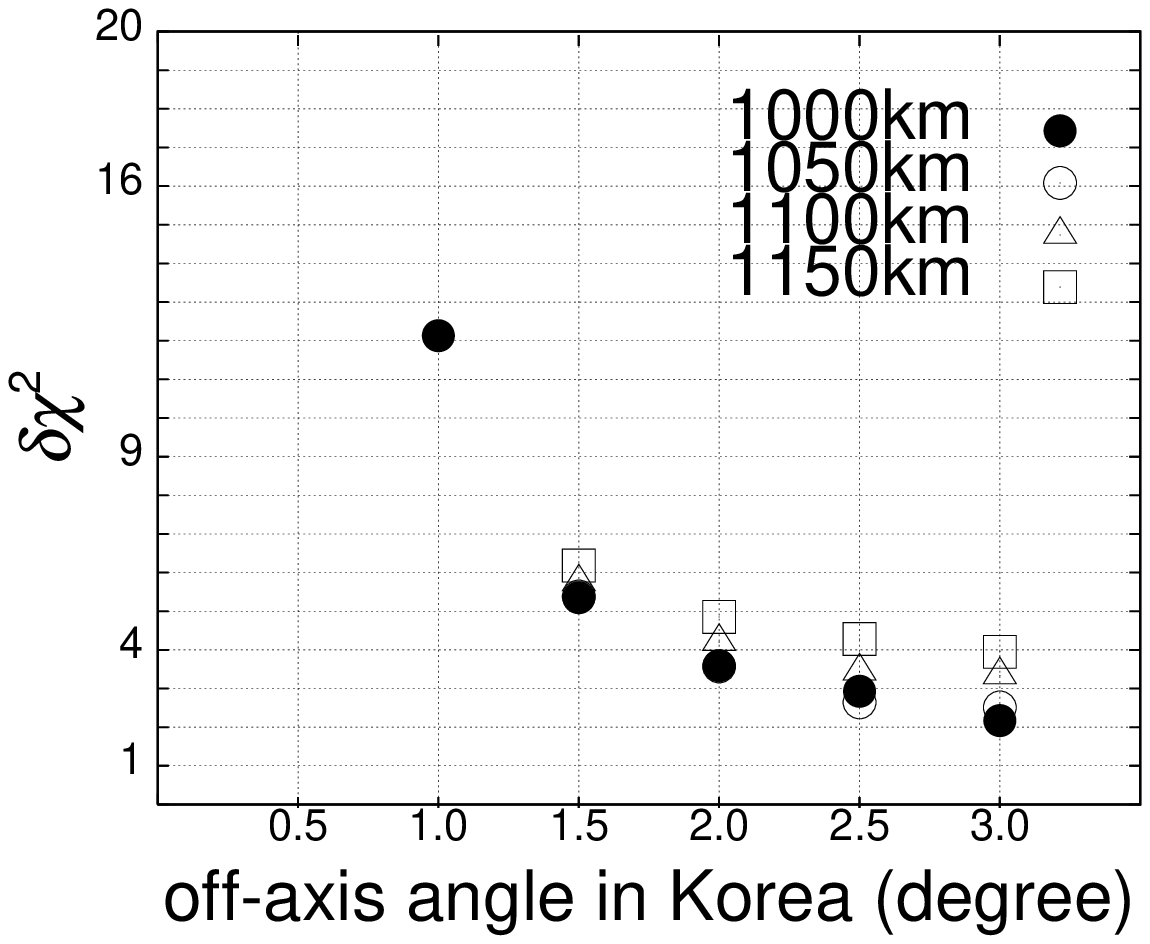}
~~~~~
\includegraphics[height=.24\textheight]{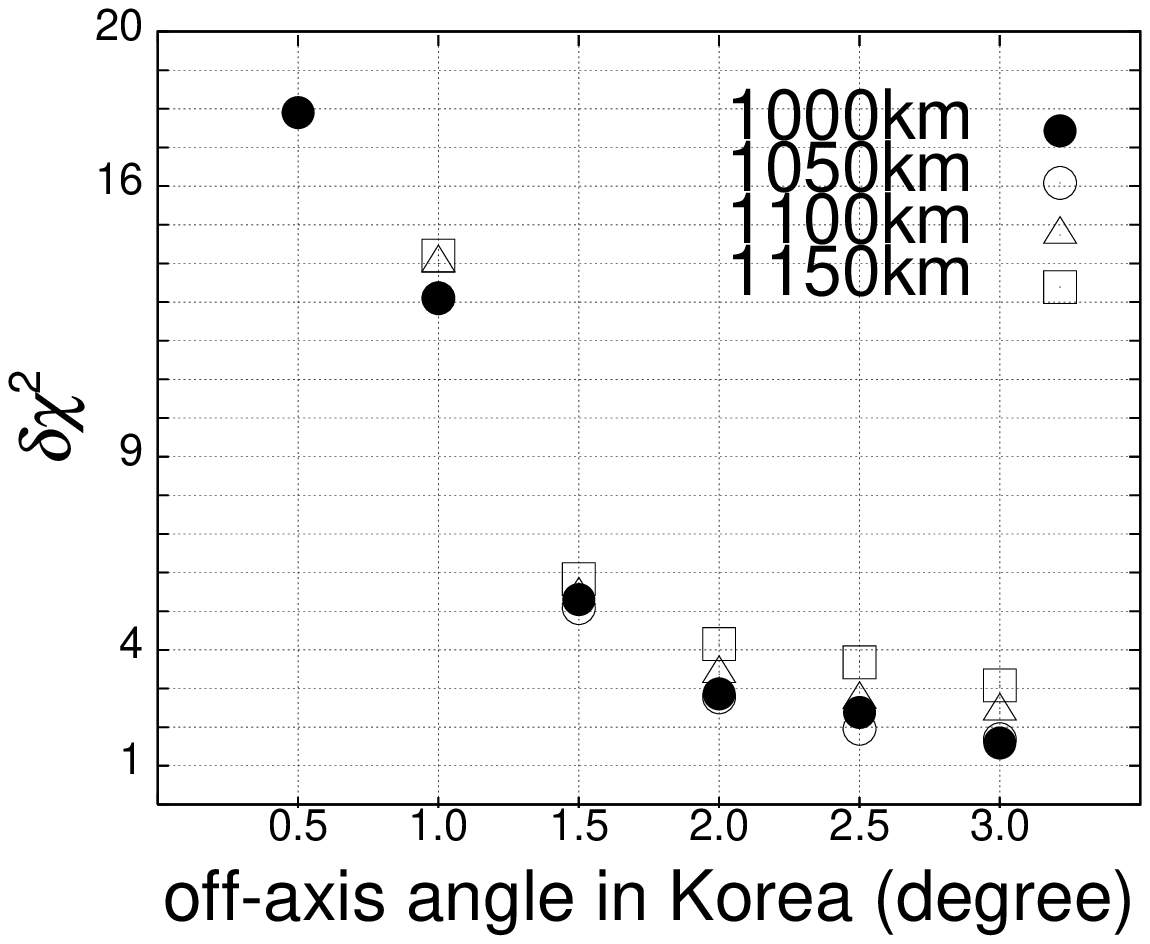}
\caption{$\delta\chi^2$ as functions of the off-axis angle and
the baseline length from J-PARC,
when the normal hierarchy 
($\dm{13}=2.5 \times 10^{-3} {\mbox eV}^2 > 0$)
with $\srct=0.10$ and $\dmns=0^\circ$
is assumed in generating the data events,
and the inverted hierarchy
($\dm{13}< 0$) is assumed in the fit. 
The left hand figure is for the $2.5^\circ$ off-axis beam at SK,
and the right hand one is for the $3.0^\circ$ beam.
}
\label{fig:hierarchy-L-angle}
\end{figure}
Since our concern is the possibility to distinguish the neutrino mass 
hierarchy and to measure $\srct$ and $\dmns$,
we study how the above ``data'',
to be gathered both at SK and a detector in Korea, 
can constrain the model parameters by using the $\chi^2$ function
\begin{equation}
\label{chi^2 define}
\chi^2 = \chi^2_{\rm SK} + \chi^2_{\rm Kr} + \chi^2_{\rm sys} 
+ \chi^2_{\rm para}\,.
\end{equation}
Here the first two terms, $\chi^2_{\rm SK}$ and $\chi^2_{\rm Kr}$,
measure the parameter dependence of the fit to the SK and the Korean 
detector data,
\begin{eqnarray}
\label{eq:chi^2event}
 \chi^2_{\rm SK,Kr}
=
\sum_{i}
\sum_{\alpha=e,\mu}
\left(
\displaystyle\frac
{(N_\alpha^{i})^{\rm fit} - (N_\alpha^{i})^{\rm true}}
{\sqrt{(N^i_\alpha)^{\rm true}}}
\right)^2\,,
\end{eqnarray}
where the summation is over all bins
up to 5.0GeV for $N_{\mu}$, 1.2GeV for $N_{e}~$at SK,
and 2.8GeV for $N_{e}~$at Korea.
We calculate $(N^i_{\mu,e})^{\rm true}$ 
by assuming the
following input (``true'') values:
\begin{equation}
\left.
\begin{array}{l}
(\dm{13})^{\rm true} = 2.5 \times 10^{-3}~\mbox{eV}^2 ~~ ( >0) \,,\\
(\dm{12})^{\rm true} = 8.3 \times 10^{-5}~\mbox{eV}^2\,, \\
(\sin^2 \theta_{\rm atm})^{\rm true} = 0.5 \,,  \\
(\ssun)^{\rm true} = 0.84 \,, \\ 
(\srct)^{\rm true} = 0.1 \,, ~~0.06 \,, \\
(\dmns)^{\rm true} = 0^{\circ},~ 90^{\circ},~ 180^{\circ},~ 270^{\circ} \,,
\end{array}
\right\}
\label{eq:input}
\end{equation}
with the constant matter density, $\rho^{\rm true}=2.8~{\rm g/cm}^3$ for T2K
and $\rho^{\rm true}=3.0~{\rm g/cm}^3$ for the Tokai-to-Korea experiments.
Note that in eq.~(\ref{eq:input}),
we assume the normal hierarchy ($\dm{13}>0$) as input (``true'')
and examine several input values of $\srct$ and $\dmns$.
The fitting event number is calculated by allowing the model
parameters to vary freely and by allowing for systematic errors.
In our analysis,
we assign $3\%$ errors :
$i$. the normalization of each neutrino flux,
$ii$. the CCQE cross sections of the $\nu_{e,\mu}$ and $\bar{\nu}_{e,\mu}$ ,
$iii$. the effective matter density along the each baseline,
to SK and to Korea,
and 
$iv$. the fiducial volume of SK and the Korean detector.
These systematics errors make the third term of eq.(\ref{chi^2 define}),
$\chi^2_{\rm sys}$.
The last term in eq.(\ref{chi^2 define}),
$\chi^2_{\rm para}$ accounts for the present constraints 
on the model parameters.
Here we interpret the $90\%$ CL lower bound on 
$\satm$ in eq.~(\ref{atm-data})
as the $1.96\sigma$ constraint from
$\sin^2 2\theta_{\rm atm} = 1 \pm 0.04$, 
and the asymmetric error for $\tan^2 \theta_{\rm sol}$ in eq.~(\ref{sol-data}) 
has been made more symmetric for $\ssun$ as $0.84\pm0.07$.
The error of each mass-squared differences is
$0.5\times 10^{-3}$eV$^2$ for $|\dm{13}|$ and
$0.6\times 10^{-5}$eV$^2$ for $\dm{12}$.
We do not include the bounds on $\srct$ in eq.~(\ref{rct-data}),
in our $\chi^2$ function.
In total, our $\chi^2$ function depends on 16 parameters,
the 6 model parameters and the 10 normalization factors.

First, 
we search for the best place for the detector in Korea and
the best combination of the off-axis angle for SK and the Korean
detector to determine the sign of $\dm{13}$.
We show in Fig.\ref{fig:hierarchy-L-angle} the $\delta \chi^2$ as 
functions of the off-axis angle and the baseline length in Korea,
when the data are generated for the normal hierarchy with
eq.~(\ref{eq:input}),
and the fit has been performed by assuming the inverted hierarchy.
We set $\sin^2 2\theta_{\rm rct}^{\rm true} = 0.10 $ and 
$\dmns^{\rm true} = 0^{\circ}$ in this analysis.
The left hand figure shows the $\delta \chi^2$ for the $2.5^{\circ}$ 
off-axis beam at SK,
and the right hand one is for the $3.0^{\circ}$ off-axis beam at SK.
The four symbols, solid circle, open circle, triangle, and square are 
for $ L =$ 1000km, 1050km, 1100km, and 1150km, respectively.
When the off-axis angle at SK is $2.5^{\circ}$,
the $0.5^{\circ}$ beam does not reach the Korean coast; see Fig.~\ref{fig:map}.
It is clear from these figures that the best discriminating power is obtained
for the combination $L$ = 1000km and $0.5^{\circ}$, 
which is available only when the off-axis angle at SK is $3.0^{\circ}$ 
(right figure).  
For this combination, we can distinguish the inverted hierarchy from 
the normal one at more than 4$\sigma$ level.
For the same baseline length,
lower off axis angle beams give better discriminating power
because the neutrino flux with smaller off-axis angle is harder 
\cite{t2k, ichikawa}, and the stronger matter effect to help us 
distinguish the neutrino mass hierarchy \cite{lipari99,barger01,t2b}.

\begin{figure}
\includegraphics[height=.20\textheight]{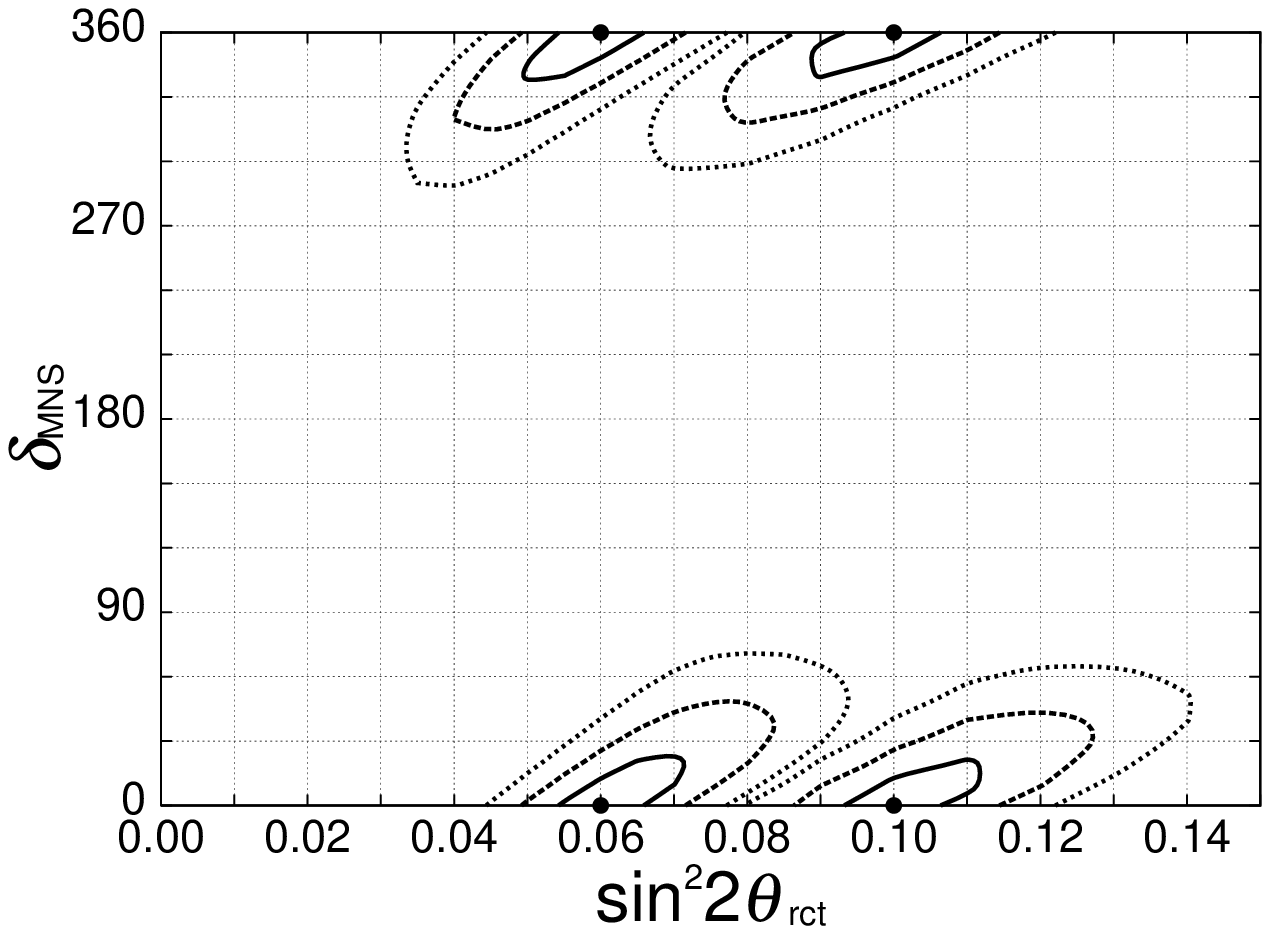}
~~~~~
\includegraphics[height=.20\textheight]{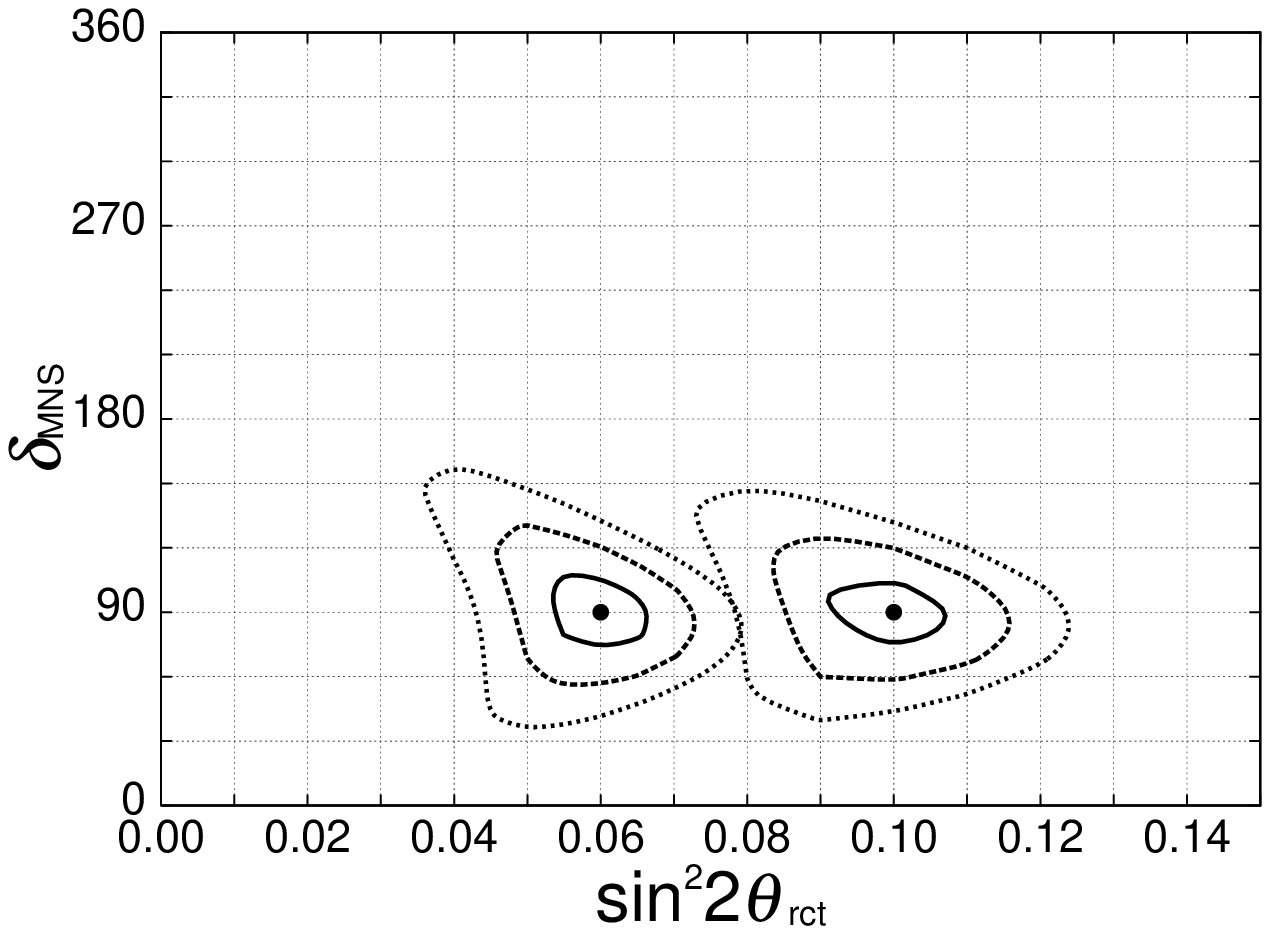}
\end{figure}
\begin{figure}
\includegraphics[height=.20\textheight]{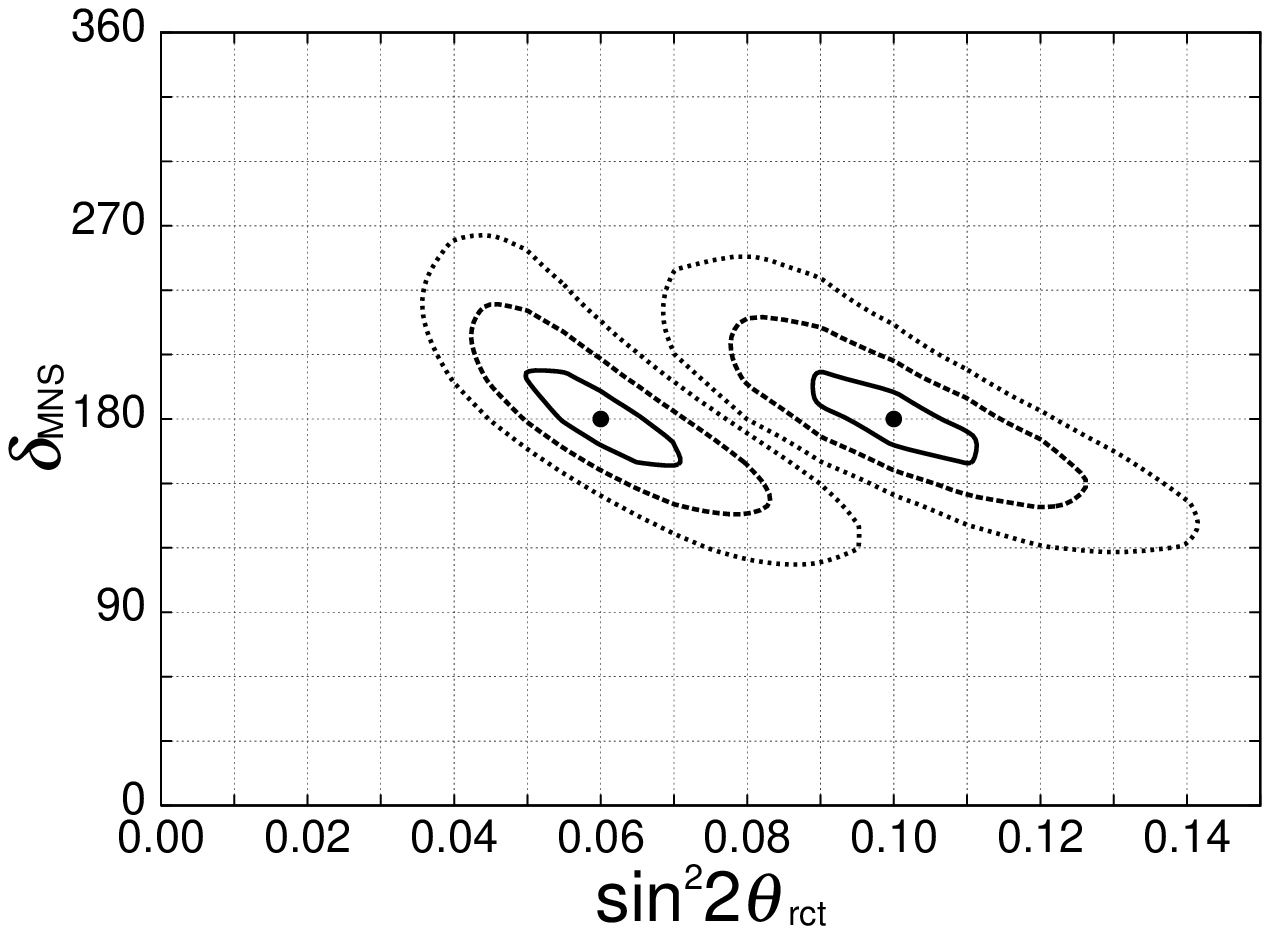}
~~~~~
\includegraphics[height=.20\textheight]{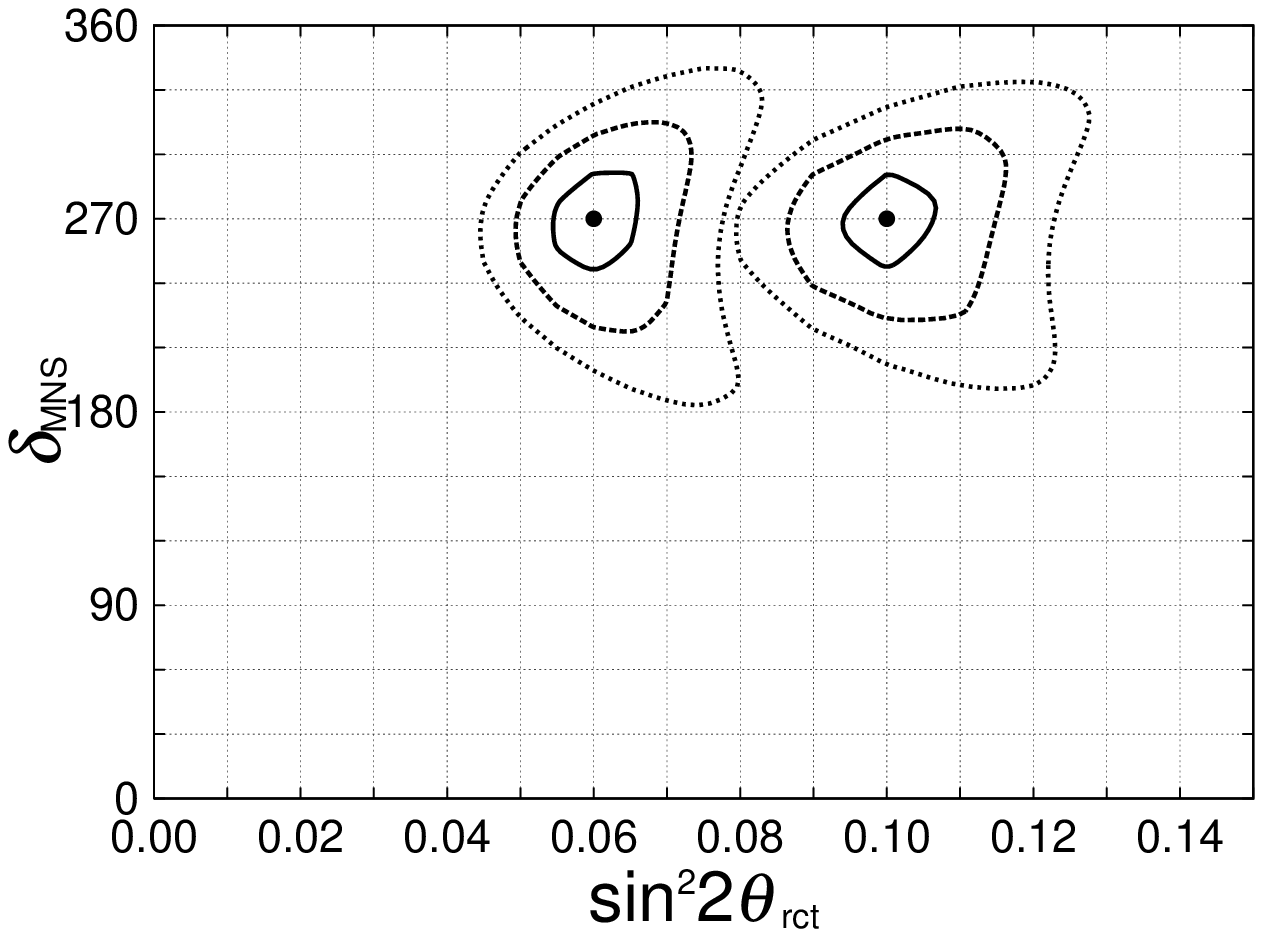}
\caption{
Allowed region in the plane of $\srct^{\rm true}$ and 
$\dmns^{\rm true}~$, when the event numbers at SK and Korea are calculated 
for the parameters of eq.~(\ref{eq:input}).
In each figure, the input points 
($\srct^{\rm true}$, $\dmns^{\rm true}$) 
are shown by solid-circles, and the regions where the 
$\delta \chi^2$ is less than 1, 4, 9 are depicted by solid, dashed and 
dotted boundaries, respectively.
}
\label{fig:cp-contr}
\end{figure}
We also examine the capability of the Tokai-to-Korea long baseline
experiments for measuring the CP phase.
We show in Fig.~\ref{fig:cp-contr} regions allowed by this experiment
in the plane of $\srct$ and $\dmns$.
The mean values of the inputs are calculated for the parameters of 
eq.~(\ref{eq:input}).
In each figure, the input points
($\sin^2 2\theta_{\rm rct}^{\rm true}$, $\dmns^{\rm true}$) are shown by
solid-circles for $\sin^2 2\theta_{\rm rct}^{\rm true} =0.10$,
and $0.06$.
The regions where the $\delta \chi^2$ is less than 1, 4, 9 are 
depicted by solid, dashed and dotted boundaries, respectively.
Even though we allow the sign of $\dm{13}$ to vary in the fit,
no solution with the inverted hierarchy that satisfy 
$\delta\chi^2 <9$ appear in the figure.  
From these figures, 
we learn that $\dmns$ can be constrained to $\pm 30^\circ$
at 1$\sigma$ level, when $\sin^22\theta_{\rm rct}^{\rm true}>0.06$.
It is remarkable that we can distinguish between 
$\dmns = 0^\circ$ and $180^\circ$,
which has been found difficult in previous studies 
\cite{koike00,barger01,aho03}.  

In this talk, we present the possibility of solving the degeneracy of
the neutrino mass hierarchy and constraining $\srct$ and $\dmns$ uniquely
by measuring the T2K off-axis beam in Korea.

\begin{theacknowledgments}
I thank our colleagues
Y.~Hayato,
A.K.~Ichikawa,
T.~Ishii,
I.~Kato,
T.~Kobayashi
and
T.~Nakaya,
learn about the K2K and T2K experiments.
\end{theacknowledgments}

\end{document}